**Quantum criticality and the formation of a putative electronic liquid crystal in $Sr_3Ru_2O_7$**

A.P. Mackenzie [1*], J.A.N. Bruin[1], R.A. Borzi [2], A.W. Rost [1,3] & S.A. Grigera [1,4]

[1] Scottish Universities Physics Alliance, School of Physics and Astronomy, University of St Andrews, St Andrews KY16 9SS, United Kingdom.

[2] Instituto de Investigaciones Fisicoquímicas Téoricas y Aplicadas, and Departamento de Física, Facultad de Ciencias Exactas, UNLP-CONICET, 1900 La Plata, Argentina.

[3] Laboratory of Solid State Physics, Department of Physics, Cornell University, Ithaca, NY 14853, USA.

[4] Instituto de Física de Líquidos y Sistemas Biológicos, UNLP-CONICET, La Plata 1900, Argentina

* To whom correspondence should be addressed at apm9@st-and.ac.uk

**Abstract**

We present a brief review of the physical properties of $Sr_3Ru_2O_7$, in which the approach to a magnetic-field-tuned quantum critical point is cut off by the formation of a novel phase with transport characteristics consistent with those of a nematic electronic liquid crystal. Our goal is to summarize the physics that led to that conclusion being drawn, describing the key experiments and discussing the theoretical approaches that have been adopted. Throughout the review we also attempt to highlight observations that are not yet understood, and to discuss the future challenges that will need to be addressed by both experiment and theory.

1. *Introduction*

The focus of this paper is the layered perovskite metal $Sr_3Ru_2O_7$, in which quasi-two-dimensional conduction takes place in Ru-O bilayers [1]. Below approximately 1K, $Sr_3Ru_2O_7$ is thought to support an electronic fluid with broken orientational symmetry, the electronic analogue of a nematic liquid crystal. The purpose of the article is to review what is known about that *nematic* phase, and how it relates to the broader phase diagram in which it appears. The bulk of the paper (sections 2-8) will be devoted to a review of experimental results, with the theoretical work discussed more briefly in section 9, and some open questions highlighted in section 10.

The experimental evidence for electronic nematicity in $Sr_3Ru_2O_7$ is summarised in Fig. 1, which shows its magnetoresistance at $T$ = 100 mK in magnetic fields up to 8.5 tesla, applied at angles θ ranging from 0˚, corresponding to field along the crystallographic *c* axis, to 90˚, when the field is applied in the *ab* plane. In each of panels 1 a) and 1 b) the resistivity is studied along orthogonal Ru-Ru bond directions (the *a* and *b* directions of a fully unfolded Brillouin zone assuming tetragonal symmetry), and the tilt is performed such that the in-plane field component aligns along one of those directions. The magnetoresistance is the same along both directions over most of this field-angle phase space, corresponding to the four-fold behaviour expected of transport in Ru-4*d* bands in a two-dimensional square planar Ru lattice. However, the magnetoresistance along the two directions differs significantly in two separate, well-defined regions (indicated by green arrows), corresponding to two-fold symmetric electrical transport [2]. The 'easy axis' is found to be perpendicular to the in-plane component of the applied field; if the in-plane field is switched from one crystal axis to the other, the easy and hard axes are exchanged, similar to behaviour seen in high purity 2D electron gases [3-6].

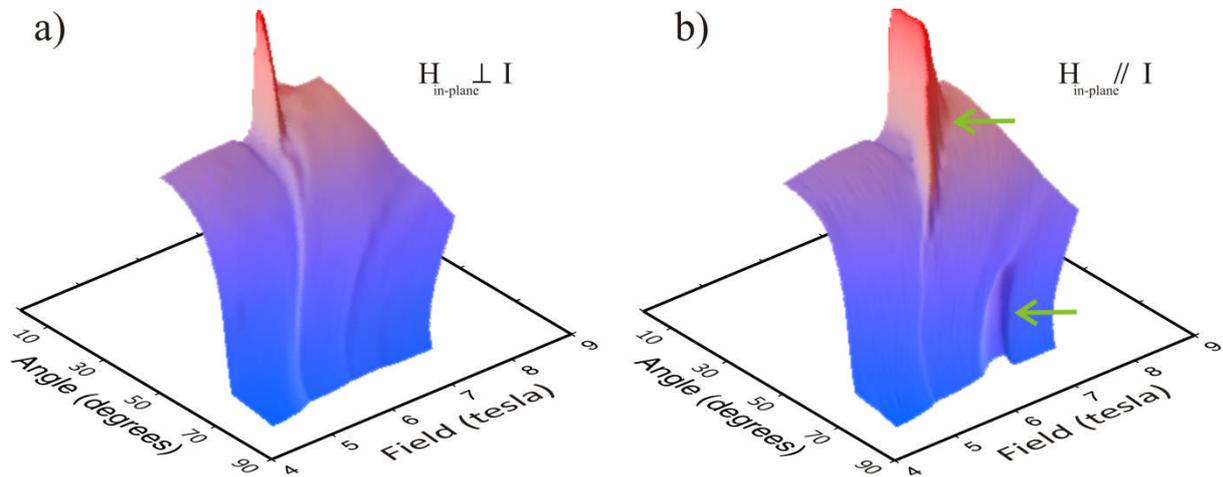

Fig. 1. Resistivity of $Sr_3Ru_2O_7$ as a function of magnetic field and tilt angle from the *ab* plane, at 100 mK. If the in-plane field component is parallel to a Ru-Ru bond direction and parallel to the applied current, the resistivity is much larger than for the orthogonal current direction in two distinct regions of the phase diagram, marked by green arrows. The tilt shown is parallel to the tetragonal *a* axis. If the field is rotated by 90 degrees in the plane, the direction of the anisotropy also switches. Figure adapted from [2].

Although the resistivity changes are striking for a metal at low temperatures, the essential physics in $Sr_3Ru_2O_7$ seems to be that of a series of electron fluids. De Haas-van Alphen oscillations can be observed in the anisotropic region for low θ [7], and searches for structures such as stripes that break translational symmetry and open a gap have not revealed anything. In any correlated electron material such as $Sr_3Ru_2O_7$, there will be a significant electron-lattice coupling. A physical property such as magnetoresistance changing from being four-fold to two-fold therefore has to be accompanied by a change in the crystal lattice. An obvious question, therefore, is whether a large structural transition might be the driver of the observed anisotropy. In the original paper reporting the magnetoresistance [2], elastic neutron scattering put an upper limit of ~ $10^{-5}$ on the lattice anisotropy change associated with entering the state supporting the strongly two-fold transport, indicating that the effect is intrinsic to the correlated electron fluid rather than being lattice-driven. Recently, an important measurement has been performed of the anisotropic thermal expansion in the phase, showing a relative lattice parameter change of 2 parts in $10^7$ for small θ, increasing to approximately 1 part in $10^6$ for 10° < θ < 15° [8].

Given the facts summarised above, the best current classification for the anisotropic in-plane behaviour of $Sr_3Ru_2O_7$ is as a nematic electronic fluid [9]. The actual anisotropy probably reflects scattering from domain walls [2], but evidence to date suggests that, whatever the details of domain formation, the underlying microscopics involve a state with orientational but not translational order. Future experiments may of course uncover translational order; if that were to happen, it would be more appropriate to describe the state as a smectic fluid, but the fact that we are dealing with a fluid with a very weak back-coupling to the lattice seems to have been established.

In the remainder of this brief review we will attempt to summarise the properties of the overall phase diagram in which this unusual behaviour appears. A key point of emphasis will be the extremely high purity that is required for the behaviour shown in Fig. 1 to be observable. For that reason, coupled with space restrictions, we will not be able to review other interesting aspects of the physics of $Sr_3Ru_2O_7$ such as the results of chemical doping experiments, the properties of the $Sr_{3-x}Ca_xRu_2O_7$ solid solution or those of eutectic mixtures of $Sr_3Ru_2O_7$ with other ruthenates. Coverage of those topics will have to await a long and comprehensive review of the field.

2. **Crystal structure, material growth and characterisation**

Building on early reports of the synthesis of $Sr_3Ru_2O_7$ [10-14], structural powder neutron scattering studies reported the crystal structure to be Bbcb [15-17], with no difference in the orthogonal in-plane lattice parameters. The reason for the symmetry being lowered from tetragonal is the presence of oxygen octahedra rotated by approximately 7˚, centred on Ru atoms whose positional shift due to this rotation was expected to be tiny. Subsequent neutron and convergent beam electron diffraction studies on single crystals reported a room temperature difference in the *a* and *b* lattice parameters to be five parts in $10^4$, meaning that the in-plane Ru atoms can be thought of as square planar to a good approximation [18].

From the first studies of the material, $Sr_3Ru_2O_7$ was known to be close to ferromagnetism. In fact, the first reported single crystals, which were flux grown in Pt crucibles [19] were reported to be itinerant ferromagnets. Once higher purity crystals were grown in an image furnace the ground state was established to be a strongly enhanced paramagnet with a Wilson ratio of approximately 10 [1] but ferromagnetism appeared to result from the application of hydrostatic pressure. Subsequent work [20-22] then revealed that uniaxial rather than purely hydrostatic pressure led to long-range magnetic order.

Although the ground state of $Sr_3Ru_2O_7$ did not prove to be ferromagnetic, a key manifestation of its proximity to ferromagnetism was the appearance of metamagnetism, namely the development of enhanced magnetism in an externally applied magnetic field [23,24]. The first generation of image furnace grown crystals had residual resistivities of several μΩcm, corresponding to mean free paths of a few hundred ångstrom. In a breakthrough that proved vital to the observation of the nematic physics, Perry and Maeno deduced that in-plane Ru vacancies were the most likely source of the remaining defect scattering, and improved the low temperature mean free path by approximately an order of magnitude [25].

### 3. Electronic structure

$Sr_3Ru_2O_7$ is the $n$ = 2 member of the Ruddlesden-Popper series $Sr_{n+1}Ru_nO_{3n+1}$, in all of whose members Ru exists in the formal valence of 4+. The electronic structure of the materials is based on partially filled bands formed from the hybridisation of Ru $4d$ levels with oxygen $2p$ states. The crudest approximation to the electronic structure comes from assuming a tetragonal crystal structure and a simple bilayer split [26]. However, the true Bbcb structure leads to a √2 x √2 reconstruction of the in-plane Brillouin zone, and a complicated Fermi surface involving hybridising bands with Ru $4d_{xy}$, $4d_{xz}$, $4d_{yz}$ and even $4d_{x^2-y^2}$ character [27]. Including spin-orbit coupling is also important; this has been done both in LDA calculations [28] and in tight-binding approaches that have been constructed to form the foundations for extended many-body calculations [29-32]. Experimentally, several studies have been performed on $Sr_3Ru_2O_7$ using Angle Resolved Photoemission (ARPES) [28, 33, 34], resulting in the Fermi surface shown in Fig. 2 and described in its caption. Strong many-body mass enhancements are evident both from the ARPES band dispersions and from measurements of the masses in dHvA experiments [7, 35, 36]. These are particularly pronounced in $γ_3$, which is believed to be in the vicinity of a van Hove singularity [37], a point which is stressed in a number of the theoretical models reviewed in section 9 of this article.

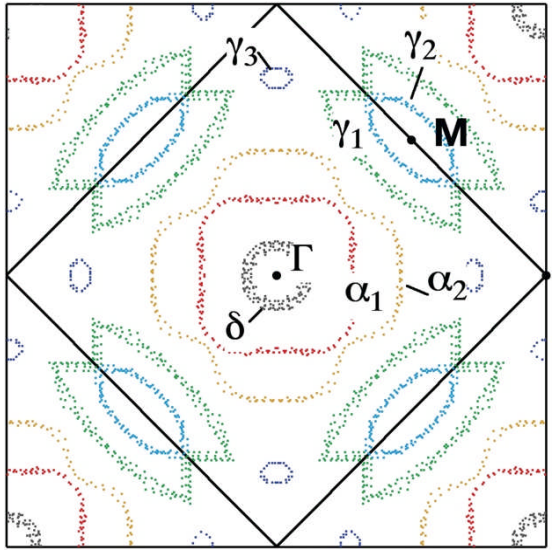

Fig. 2. The Fermi surface of $Sr_3Ru_2O_7$ determined by the angle-resolved photoemission study of Tamai and co-workers [28]. The $α_1$ and $α_2$ sheets have strong Ru $4d_{xz,yz}$ character, while the δ sheet can be identified with the $4_{x^2-y^2}$ levels (all hybridising with oxygen $2p$ levels). The $γ_1$, $γ_2$ and $γ_3$ sheets are hybridised mixtures of the Ru $4d_{xy}$, $4d_{xz}$ and $4d_{yz}$ levels. $γ_3$ is associated with a very narrow band in the vicinity of a van Hove singularity. The bilayer splitting is evident from inspection of $α_1$ and $α_2$, but it is not clear whether weak splitting exists for one or more of the other Fermi surface sheets and remains unresolved by the ARPES experiment. Bilayer-bilayer coupling is weak; in-plane electrical conductivity is several hundred times larger than that perpendicular to the bilayers [1].

### 4. Magnetic field-angle phase diagram determination from magnetic susceptibility

Although the magnetoresistance study summarised in Fig. 1 above is a sensitive indicator of electronic nematicity, it is important that it be combined with measurements sensitive to crossing thermodynamic phase boundaries. The low temperature phase diagram of $Sr_3Ru_2O_7$ contains both magnetically coupled first order phase transitions and crossovers. Low frequency a.c. magnetic susceptibility ($7 < f < 100$ Hz) proved to be an ideal probe of these phenomena, because crossing a first order phase boundary yields a small signal in the imaginary part of the response, which can be detected with high sensitivity. Crossovers, on the other hand, are seen via peaks in the real part [24]. The imaginary signal on crossing the first order boundary is due to hysteretic dissipation, and is accompanied by hysteresis in either the real part of the a.c. susceptibility or in a measurement of the d.c. magnetisation. Both the latter effects have been observed in $Sr_3Ru_2O_7$ [24, 38] for angles near the *ab*-plane, where the transition is strongly first order at low temperatures, but the experimental sensitivity is not as good as that with which the imaginary part of the a.c. susceptibility can be followed. An imaginary signal has also been observed in low temperature studies of the alternating-field magnetoresistance across one of the phase boundaries [39].

In Fig. 3 we show how first order phase transitions in $Sr_3Ru_2O_7$ evolve with sample purity. As seen in Fig. 3a, if the mean free path is approximately 300 Å, there is a surface of first order magnetic phase transitions terminating at approximately 1.2 K for fields applied in the *ab* plane, with the termination temperature falling to below 100 mK when the field is rotated to within 10° of the *c* axis. Increasing the mean free path by an order of magnitude results in a qualitative change (Fig. 3b) [40,41]. The behaviour for fields applied near the *ab* plane is similar, but as the field is rotated towards the *c* axis, the first order surface splits into two at θ ≈ 30˚ from the *c* axis, and their termination temperatures rise.

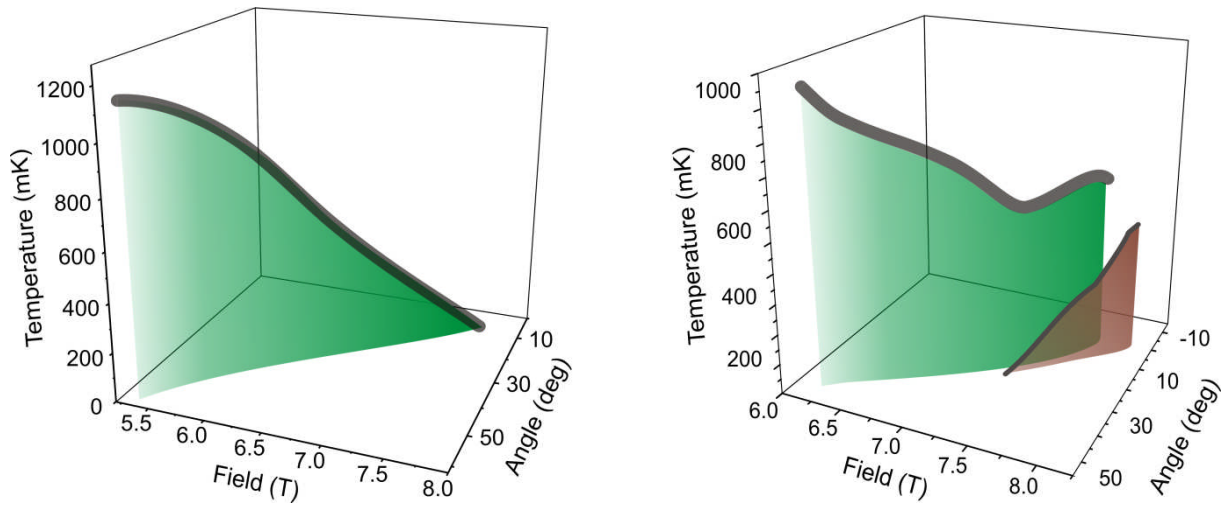

Fig. 3. a) The green surface shows the locus of points in the (*T*,θ,*H*) phase diagram at which a peak is seen in the imaginary part of a low frequency a.c. magnetic susceptibility signal, in crystals with mean free path of approximately 300 Å. It corresponds to a surface of first order phase transitions. The thick black line traces the overall maximum seen in the real part of the signal in the same measurements [24]. Panel b) shows an identical experiment performed on crystals with mean free path approximately 3000 Å, revealing the appearance of a second surface of first order transitions (brown) [41]. In both panels, the tilt shown is parallel to the tetragonal *a* axis.

## 5. Relationship between transport anisotropy and metamagnetism

A summary of the experimentally determined low temperature field-angle phase diagram is shown in Fig. 4, in which the loci of prominent metamagnetic crossovers are marked along with those of phase transitions. As can be seen from inspection of Figs. 1 and 3, there is a clear correlation between the metamagnetic phase transitions and crossovers and the changes from four-fold to two-fold transport signatures. In some parts of the phase diagram, this correlation is precise. For example, the lower field transition of Fig. 3b coincides, within experimental resolution, with the onset of the anomalous transport. However, for small angles from the *c* axis (less than 12°), strong anisotropy extends slightly beyond the field of the upper first order phase transition, something that is currently the subject of experimental investigation [42].

The region of resistive anisotropy for θ < 35˚ (green shading) has been investigated more thoroughly than the one for θ > 55˚ (blue shading). So far, there is no evidence that the magnetic signals that mark the entry to, or exit from, the blue region of resistive anisotropy are phase transitions. However, this inference is based primarily on the non-observation of dissipation or hysteresis, and it is possible that these are in fact first order phase transitions whose hysteretic signal falls below current experimental resolution.

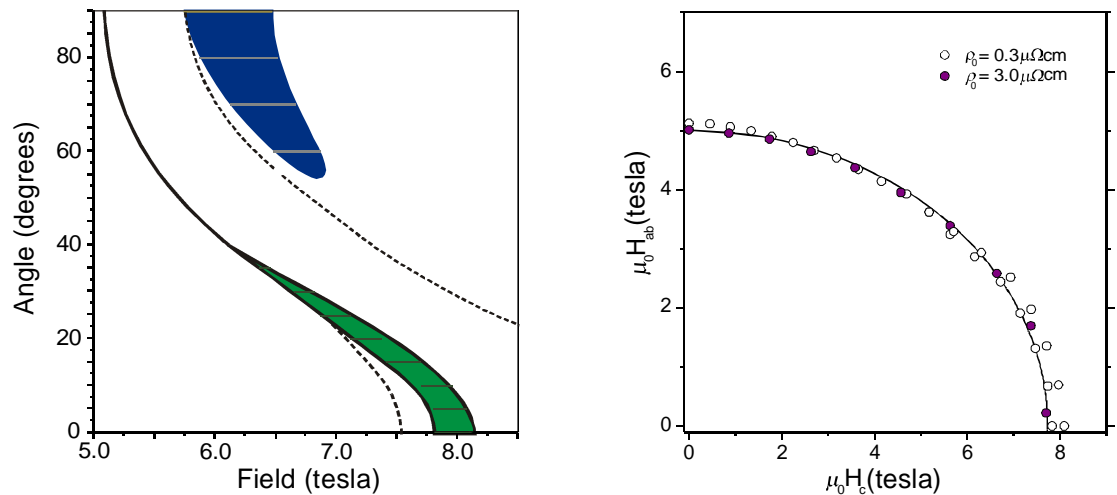

Fig. 4. a) Summary of regions of anisotropic transport (green, blue), first order metamagnetic transitions (solid black) and crossovers (black dashed), as a function of applied field magnitude and angle with respect to the crystal c axis. Recent results [42] indicate that the region shaded in green extends somewhat beyond the second critical field for angles below 12°. A further metamagnetic feature has also been identified near 13 T for low angles using torque magnetometry [43]. b) Summary of the principal metamagnetic transition for low purity (coloured circles) and high purity (white circles) samples as a function of the components of the applied field in the *ab* plane and *c* axis. In both panels, data were taken with the field tilted parallel to the tetragonal *a* axis, with $T$ = 100 mK.

A second key point about the relationship between magnetism and nematicity is that, independent of whether one is entering or leaving a nematic part of the phase diagram, the magnetic transition or crossover involves an increase of moment with applied field. This is illustrated in Fig. 5, and described in more detail in its caption.

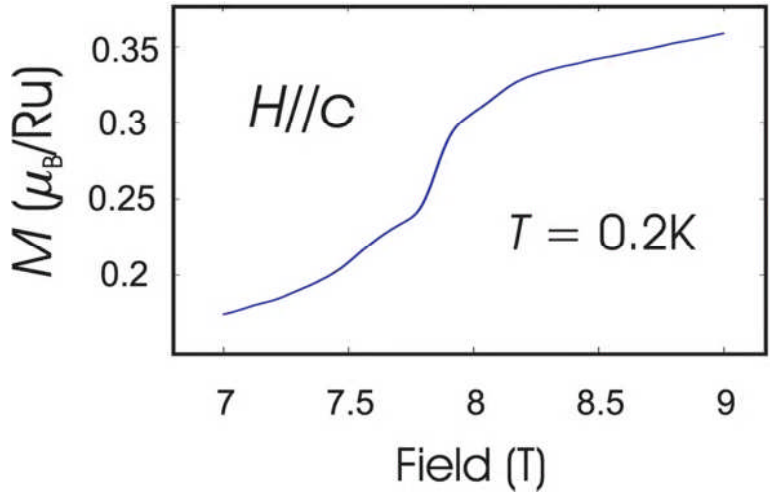

Fig. 5.  Magnetisation on a high purity single crystal of $Sr_3Ru_2O_7$ measured in a capacitive Faraday magnetometer at 200 mK.  Three consecutive rises in magnetic moment are seen: a crossover at 7.5 T is followed by a sharp rise at 7.8 T and a third, smaller rise at 8.1 T.  The two higher field features show a dissipative component in a.c. susceptibility measurements (see for example [40]).  Graph from [44].

## 6.  Probing nematicity through transport anisotropy

A central issue when studying electronic nematicity is whether it can be observed in the absence of a symmetry-breaking field, when the preferred direction is not fixed to either of the orthogonal axes in a square lattice, and domain formation is likely. In 2D electron gases, the symmetry appears to be broken by some macroscopic feature of the devices, though its origin remains unknown [9].  In $Sr_3Ru_2O_7$, no in-plane anisotropy is observed when the magnetic field is applied along the high-symmetry crystallographic $c$ direction. However, in the region shaded green in Fig. 4, anisotropy can be induced by tilts away from $c$ of as little as 1˚ [42].  A second important point is that although no anisotropy is seen for $H$ precisely parallel to $c$, a large effect is still observed in all components of the resistivity tensor on entering the phase. The resistivity jumps by a factor of approximately two, while the Hall component shows a pronounced dip [45].

The most natural explanation for these observations is that a microscopic phase transition to a nematic state takes place even when the field is parallel to $c$, but that domain formation of some sort masks the observation of a macroscopic anisotropy until an in-plane field component unbalances either the domain orientation or occupation.  Since it couples in this way, the in-plane field must be conjugate to the microscopic order parameter, so is expected to round out any continuous thermal transition to the nematic state.

Although the above explanation seems to be both natural and self-consistent, it is based on assumptions that are not fully proven.  What has actually been measured in $Sr_3Ru_2O_7$ is a nematic susceptibility, which is observed to be very small across most of the $H$-θ plane and very large in certain well-defined regions of it, as summarised in Fig. 4.  The domains that are postulated to exist for $H \;//\; c$ have not been directly observed, but an analysis of the damping of de Haas-van Alphen oscillations suggests that they would be associated with a characteristic length scale of approximately 5000 Å [7].

### 7. Determination of thermodynamic phase boundaries

The parameter space associated with the *H*-θ phase diagram of Fig. 4 is large, and the extension to finite temperatures is not yet complete. However, considerable work has been done for θ = 0° (*H* // *c*). A summary of the findings of a number of experiments is given in Fig. 6 [46]. The curvature of the two highlighted first-order transition lines is unusual, because the Clausius-Clapeyron relationship implies that the entropy is higher in the region between the lines than outside them.

This issue was investigated by a combined study of the magneto-caloric effect and specific heat [44,47] from which several important deductions could be made. Firstly, the identification of the purple triangles from Fig. 6 as first order transitions between equilibrium phases was confirmed by a quantitative comparison of entropy changes deduced independently from direct measurement and the application of the Clausius-Clapeyron law. This also confirmed the striking increase of entropy on entry to the bounded region of Fig. 6. Secondly, temperature dependent specific heat studies [44,47] provided evidence, consistent with that from thermal expansion measurements [8], that the 'roof' of the phase is a line of second order phase transitions.

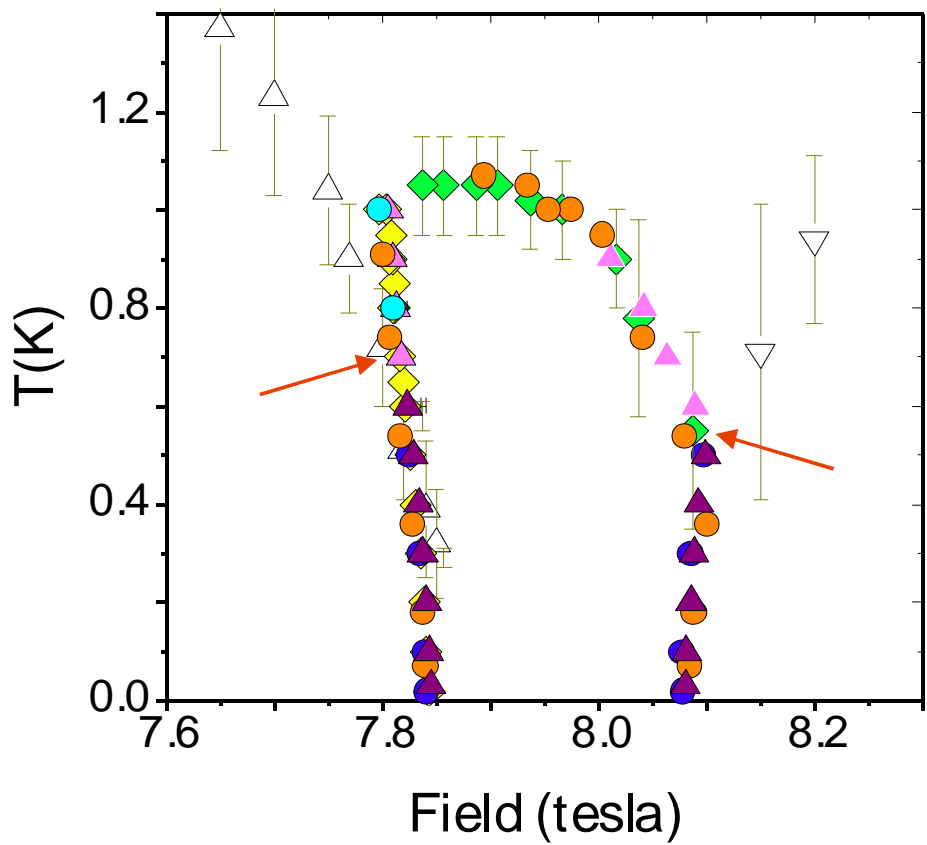

Fig. 6. Empirically determined phase boundaries from a.c. susceptibility (coloured triangles), resistivity (yellow and green diamonds), magnetostriction (blue circles), combined with crossovers

determined from maxima and minima of thermal expansion (white triangles). The red arrows mark the maximum temperatures at which hysteresis is seen in a.c. susceptibility [46].

As discussed above, there is circumstantial evidence for the existence of nematic domains within the phase for this field orientation, and various authors speculated that the extra entropy might be associated with these. However, the value of the entropy jump (≈10 mJ/Ru-mol K or 0.1% of R ln(2)/Ru-mol at 400 mK) suggests another source, and recent thermal expansion data seems to show that the unusual slope of the phase boundaries continues to be seen at higher θ, for which the domain density is inferred to be much weaker than for $H // c$ (Fig. 7). The strange entropic properties of the phase seem, therefore, to be an intrinsic feature of it, and a key challenge for the construction of a comprehensive theory of the properties of $Sr_3Ru_2O_7$.

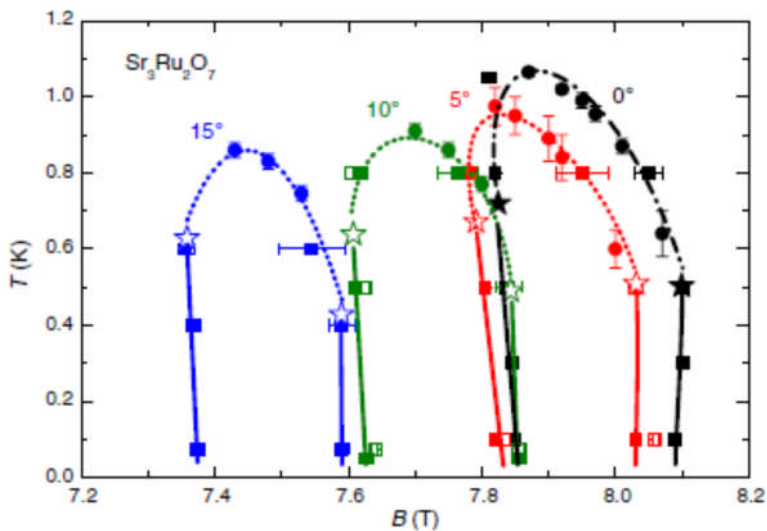

Fig. 7. The boundaries of the nematic phase at different tilt angles. The squares and circles denote features in the magnetorestriction and thermal expansion respectively [8].

A second particularly notable feature of the phase is the disorder criterion for its observation. Using transport data and the renormalized band parameters obtained from ARPES and dHvA [7, 28, 35, 36], it is possible to estimate the critical scattering rate above which the phase is no longer visible. That rate is of order $T_c$, which is a characteristic signature of microscopic physics involving an anisotropic distortion in $k$ space [9]. A 'd-wave Pomeranchuk distortion' (a four-fold symmetric Fermi surface sheet distorting spontaneously to a two-fold symmetric one) is entirely consistent with this disorder sensitivity [46]. That does not imply that it is the only possibility, but it is the one studied in detail in most of the theories reviewed in section 9.

## 8. Signatures of quantum criticality in the broader $Sr_3Ru_2O_7$ phase diagram

Since this review is primarily concerned with electronic liquid crystals, the order of presentation has been chosen to emphasise the evidence for the electronic nematic state in $Sr_3Ru_2O_7$. However, the material first attracted widespread attention due to the observation of a series of signatures suggesting that its overall phase diagram contained one or more quantum critical points [23, 48], and a number of experiments suggest that the quantum criticality and nematic phase formation are intimately linked.

The first signs of quantum critical behaviour were seen in magnetotransport, specifically the field dependence of the exponent, α, of the temperature dependent part of the electrical resistivity $\rho = \rho_o + AT^\alpha$ [23, 48]. For low temperatures and fields either lower or higher than the critical field $\mu_0 H_c \approx 7.9$ T, α = 2, as expected in a Fermi liquid. As $H_c$ is approached, the crossover temperature to the Fermi liquid falls smoothly, as evidenced both by the change in α and the field dependence of a peak in the specific heat associated with a thermodynamic crossover scale [47]. At $H_c$, α = 1 to high accuracy between 20K and the onset of the nematic phase at ~ 1K [42,47]. Quantum critical signatures have also been seen in the temperature dependences of a number of other properties, notably the electronic specific heat [23,47], the NMR relaxation rate [49] and thermal expansion [50], and the specific heat has been shown to depend on the chemical potential in a way that cannot be understood from simple band theory assumptions [51].

The second feature that is intuitively expected of a quantum critical phase diagram with high and low field Fermi liquids is a low temperature mass divergence on the approach to the critical point. This is expected because the lower the crossover temperature from the strongly fluctuating quantum critical state to the Fermi liquid, the more rapid the drop of entropy must be as $T$ decreases in that Fermi liquid. A strong peak is observed in both the specific heat and magneto-caloric effect [44], as expected on general scaling grounds [52], but the strength of the divergence is surprising. In contrast to predictions for a metamagnetic quantum critical point [53], it is observed to grow as $[(H - H_c)/H_c]^{-1}$ [44]. Some care is required in interpreting this result because the underlying electronic structure should be taken into account as well as quantum critical fluctuations [54], but it stimulates further theoretical work.

An important question is whether the behaviour of the specific heat reflects a divergence of the mass of Landau quasiparticles in the low temperature Fermi liquids above and below the critical field. Several studies have been reported of the field dependence of quantum oscillations in $Sr_3Ru_2O_7$. In the first, an apparent divergence of the quasiparticle masses of at least two of the Fermi surface sheets was reported [35], but subsequent studies performed at higher signal-to-noise ratios showed that this was not the case [36]. The origin of the divergence seen in the specific heat and magnetocaloric signals is therefore the subject of continuing investigation.

Empirically, it seems that the nematic phase is thermodynamically stabilised as part of a process that cuts off the electronic specific heat divergence, in a process that is similar to other examples of phase formation in the vicinity of quantum critical points [47, 55, 56]. Approaching from the low field side, the divergence is cut off at the first of the three magnetic moment jumps shown in Fig. 5, and the nematic phase onsets at the second of them.

The unusual entropic behaviour within the phase seems to be related to the way that it is entered from high temperature. It does not form out of an intermediate Fermi liquid, but directly from the quantum critical state. The $T\ln T$ electronic specific heat of that state is cut off at $T_c$, but some of the quantum critical degrees of freedom seem to be left in the ordered phase itself [47]. Further evidence that quantum criticality and phase formation are intimately linked comes from the observation that when the magnetic field is applied purely in the $ab$ plane, the electronic specific heat again varies as $T\ln T$ on cooling above the second 'nematic' region at ~ 6 T [38].

These observations suggest that, in, $Sr_3Ru_2O_7$, the formation of the electronic nematic is intimately linked to the physics of quantum criticality. It therefore seems likely that the influence of quantum critical degrees of freedom must ultimately be included in any theoretical approaches to the problem.

### 9. Theoretical approaches to the physics of $Sr_3Ru_2O_7$

Although the primary motivation for this article was to review the experimental progress that has been made, in this section we will address the classes of theoretical approach that have been adopted to account for the physics of $Sr_3Ru_2O_7$. The discussion will be brief, both for reasons of space and because, in the absence of a complete experimental determination of the phase diagram, theoretical understanding to date is necessarily incomplete. Nevertheless, the theoretical work so far has resulted in a productive interchange with the experimental groups, and the progress that has been made has been driven by the combination of experiment and theory. More comprehensive and broader reviews of the theory of electronic nematics and quantum criticality can be found in [9, 57]; here we focus on work directly addressed to the physics of $Sr_3Ru_2O_7$.

We first cover the existing theoretical work on quantum criticality and related phenomena. A Hertz-Millis treatment of the metamagnetic quantum critical end-point [53] closely followed the experimental work reporting its likely existence in $Sr_3Ru_2O_7$ [48]. The scattering enhancement by fluctuations was studied in the same class of model [58,59]. The bifurcation of a sheet of first order metamagnetic transitions, leading to a dislocated quantum critical point, was modelled by introducing a coupled order parameter to the Landau free energy expansion around a metamagnetic critical point [41]. The potential role of an underlying van Hove singularity (vHs) in producing a phase diagram similar to that observed in $Sr_3Ru_2O_7$ was also identified [60] and followed up in detail by others [61, 62]. The entropy measurements discussed in sections 7 and 8 have also stimulated interest from the large theoretical community working on the 'AdS-CFT' correspondence which may open a new route to calculations of certain types of strongly coupled quantum critical phenomena [63, 64].

The potential existence of nematic phases in correlated electron systems was highlighted in the 1990s and early 2000s [65-68], and considerable work done on the subject in relation to the experiments on high purity 2D electron gases. Following the suggestion (not at that point backed up with evidence of anisotropic transport) that $Sr_3Ru_2O_7$ might undergo a Pomeranchuk-type distortion of the Fermi surface [46], attention began to focus directly on this possibility [69]. Since the discovery of the anisotropy, a number of groups have worked intensively on understanding the phase diagram. A considerable body of work stresses the importance of the role played by a 2D vHs [31, 32, 68-78]. Most of that work has concentrated on single-layer systems, but the extension to bilayers has also been considered [79, 80].

In some cases, it has proved convenient to model the tendency to nematicity by inserting an explicit term in the Hamiltonian, and studying the consequences [68-75, 79], while in others, tight-binding models to the underlying band structure have been used as the starting point [32, 78]. A second class of tight-binding model has incorporated $d_{xy}$, $d_{xz}$ and $d_{yz}$ orbitals, allowing for the possibility that the nematicity might originate from the 1-D character $d_{xz}$ and $d_{yz}$ bands and the physics of a 1D vHs [29, 30, 32, 81, 82]. This would be equivalent to a form of orbital ordering, and spin-orbit coupling, which is incorporated into most of these models, can be expected to play a significant role.

Although most theories of the nematic phase have considered it as a Pomeranchuk-like Fermi surface distortion, other possibilities, such as an inhomogeneous charge state [83], a spiral spin state [76, 77] or the formation of Condon domains [84] have also been explored. In a Condon domain picture the apparent phase boundaries would have a dependence on the demagnetisation factor of the crystals, but changing the demagnetisation factor experimentally from approximately 1 to 7 did not produce the predicted effect [2]. In the absence of microscopic information about the phase, the other scenarios cannot at this stage be ruled out.

Key experimental observations that have not been addressed explicitly so far are the scale of the resistive anisotropy (approximately a factor of two) and its relationship to the behaviour of the Hall voltage, and the unusual entropic behaviour of the phase. A large change in resistivity at 1K is not easy to model, especially when the lattice parameter anisotropy is known to be tiny. It seems likely that domain wall scattering of some kind must play a role, and that the walls must scatter strongly. Although there has been one report discussing domain formation [74], no calculation of the bulk resistivity has been attempted, and for the other microscopic proposals for the nature of the phase, no estimate has been given for the expected scale of the resistive anisotropy. The intriguing entropic behaviour of the phase strongly suggests that any successful theory will have to go beyond mean-field. Very recently, calculations attempting to do that have been reported [85, 86]; we believe that more work in this direction will be required.

## 10. Open questions and future work

As we hope we have described in the bulk of this article, much has now been understood about the properties of $Sr_3Ru_2O_7$. However, a series of fundamental questions remain, both about the anisotropic phase and the quantum critical metal from which it forms. Of particular importance is the nature of the quantum critical fluctuations. Early work suggested that these could be attributed to simple Stoner itinerant metamagnetism [23, 48], but more recent observations call this into question. In particular, a detailed study has been performed of the effect of hydrostatic pressure on the metamagnetism of clean samples for fields applied in the *ab* plane. Although the end-point of the first order metamagnetic transition at 5.2 T (see Fig. 5) can be tuned to *T* < 75 mK, there are no signs of a divergence of the static magnetic susceptibility [87]. Similar questions had previously been raised following studies of the angle-dependent susceptibility of less pure compounds, though in that case the analysis had been complicated by a strongly frequency dependent signal [24]. These experiments suggest that the underlying quantum critical point is itself related to a more complex state than a simple metamagnet. Related questions concern the origin of the 'metamagnetic crossover' seen at 7.5 T for *H // c*, and of the region just above the transition at 8.1 T, in which a large nematic susceptibility is seen. These issues merit further work.

A second central challenge will be to determine the microscopic order parameter associated with the phase. Here two probes are likely to play a major role, probably in combination. Investigation of *k*-space structure is best carried out by neutron scattering, because it is suited to work at high fields and has the desired sensitivity to magnetism. Both elastic and inelastic neutron work have been performed on $Sr_3Ru_2O_7$ [88-91] but much more still needs to be done, probably in combination with resonant x-ray scattering. An ideal complement to these *k*-space studies will be spectroscopic scanning tunnelling microscopy (STM), which is a sensitive probe both of orbital order and real space texturing. One of the major questions concerning spectroscopic STM is whether a given compound cleaves properly and yields bulk-sensitive information. Ruthenates are well suited to such work because they contain charge neutral planes, and work by two groups has shown that both atomic resolution and good spectroscopic data are attainable [37, 92]. To date, however, no work has been reported at the relevant combination of sample purity, temperature and magnetic field. It is desirable that such experiments be performed, along with further careful studies of the transport characteristics in and around the phase. It is hoped that a unified understanding of these issues will also shed light on two other prominent puzzles, namely the entropic behaviour within the phase and the microscopic origin of the 'mass divergence' seen in entropy and specific heat measurements.

## 11. Conclusion

In this review we have attempted to highlight the status of research on $Sr_3Ru_2O_7$ as it stands at the end of 2011. It is necessarily a snap-shot of the current situation, because many things remain to be understood in depth. All of these are challenges not for theory or experiment in isolation, but strongly advocate for continuation of the combined approach that has led to most of the progress so far. We anticipate several more years of active research on this fascinating material.

**Acknowledgements**

We thank the collaborators not directly involved in the writing of this article for their insights and input to the field. We also acknowledge the support of the UK Engineering and Physical Sciences Research Council and the Royal Society. One of us (APM) benefits from a Royal Society – Wolfson Research Merit Award.